\documentstyle[prc,epsf,aps,twocolumn]{revtex}

\begin{document}
\title{Tensor correlations in nuclei and exlusive electron scattering} 
\author{J. Ryckebusch \footnote{E-mail :
jan.ryckebusch@rug.ac.be}, S. Janssen, W. Van Nespen and D. Debruyne}
\address{Department of Subatomic and Radiation Physics \protect\\
University of Gent, Proeftuinstraat 86, B-9000 Gent, Belgium}

\date{\today}
\maketitle
\begin{abstract}
The effect of tensor nucleon-nucleon correlations upon exclusive and
semi-exclusive electronuclear reactions is studied.  Differential
cross sections for the semi-exclusive $^{16}$O($e,e'p$) and exclusive
$^{16}$O($e,e'pn$) processes are computed by explicitly evaluating the
dynamical electromagnetic coupling to a tensor correlated nucleon
pair. In both reaction channels the tensor correlations contribute in
a very substantial way.  Tensor correlations are found to generate
more electronuclear strength than central Jastrow correlations do.

\end{abstract}


In the history of nuclear physics, it has been notoriously difficult
to detect signals that directly point towards phenomena beyond the
scope of the effective mean-field theories.  This holds in particular
for the short-range correlations that reflect the remnants of the
hard-core part of the nucleon-nucleon ($NN$) force in the medium.
Recently, manifestations for strongly correlated proton pairs and the
existence of Jastrow-like correlations emerged from the simultaneous
detection of two protons upon absorption of one (virtual) photon by an
atomic nucleus \cite{gercoprl2,blom98}.  Nuclear many-body theories
have produced vastly different predictions for the short-range
behaviour of nuclei.  The ongoing exclusive $A(e,e'pp)$ studies are
expected to provide stringent tests of these theories.  The
correlations probed in proton-proton knockout are predominantly the
state-independent scalar ones (often referred to as Jastrow
correlations) related to the hard core part of the $NN$ force.  The
tensor force, which is operative at intermediate internucleon
distances ($\approx$ 1-2.5~fm), is established to be an important
ingredient of the $NN$ force in the medium and is believed to be
another source of important $NN$ correlations which go beyond the
mean-field level \cite{benhar93}.  High momentum components in the
deuteron wave function, for example, are ascribed to a $D$-state
admixture and are a direct manifestation of the presence of tensor
correlations in the proton-neutron system.  Since long, the tensor
interaction has been established to be a rather weak but essential
ingredient of the effective $NN$ force.  Despite intensive research a
recent review \cite{fayache97} quoted its role as ``elusive''.
Earlier studies of the role played by tensor correlations in electron
scattering concentrated on inclusive $A(e,e')$ response functions for
which there are many competing effects and unambiguous information on
the tensor correlations might be difficult to extract
\cite{orlandini83,fabrocini97}.  In the near future, exclusive
experiments that aim at probing both the proton-proton and
proton-neutron correlations will be performed at MAMI and TJNAF.  At
MAMI, where central short-range correlations in nuclei are being
studied with the aid of the $A(e,e'pp)$ reaction
\cite{blom98,guenther}, high-resolution $A(e,e'pn)$ measurements have
been scheduled for the target nuclei $^3$He \cite{eddyprop} and
$^{16}$O \cite{peterprop}.  These measurements will be performed at
four-momentum transfers of the order $Q^2 \approx 0.05~(GeV/c)^2$. At
TJNAF, on the other hand, the small distance structure of nuclei will
be probed with the aid of the $^{12}$C($e,e'pN)\; (N=p,n)$ reaction at
$Q^2 \geq 2~(GeV/c)^2$ \cite{piasetzky}.  In the light of the upcoming
experiments, we present model calculations that aim at exploring the
possibility of using exclusive and semi-exclusive electronuclear
reactions to elucidate the role of tensor correlations in the {\em 
nuclear medium}.

To establish a connection between the $NN$ correlations
and the measured electronuclear cross sections, we remark that the
response of the target nucleus (with ``correlated'' ground-state wave
function $\left| \overline{\Psi_i} \right>$) to the electromagnetic probe is
determined by matrix elements of the form
\begin{equation}
\frac {\left< \overline{\Psi_f}\mid J_{\mu}(q) \mid
 \overline{\Psi_i} \right>}
{\left< \overline{\Psi_f} \mid  
\overline{\Psi_f} \right> \left< \overline{\Psi_i} \mid
 \overline{\Psi_i}  \right>}
 \; .
\label{eq:transim}
\end{equation}
We introduce ``correlated'' nuclear wave functions in a standard
manner 
\begin{equation}
\left| \overline {\Psi} \right> = \widehat{{\cal S}} \left[\prod _{i<j=1} ^A
\left( 1 - g_c(r_{ij})+f_{t\tau}(r_{ij})\widehat{S_{ij}} \vec{\tau}_i
. \vec{\tau}_j \right) \right] \left| \Psi \right> \; ,
\end{equation}
where $\left| \Psi \right>$ is the uncorrelated and normalized (Slater
determinant) wave function obtained from a mean-field calculation,
$\widehat{{\cal S}}$ the symmetrizer, $g_c$ the central correlation
function that accounts for (state-independent or Jastrow) short-range
effects and $\widehat{S_{ij}}$ the tensor operator that introduces
state-dependent correlations for the $S=1$ components in the nuclear
wave functions.  Strictly speaking the correlation operator also
contains spin-isospin and spin-orbit terms.  The $g_c$ and $f_{t\tau}$
are, however, the most important ones and without their presence
finite nuclei would simply be unbound \cite{benhar93}.  A striking
feature of the correlation functions $g_c$ and $f_{t\tau}$ is that
they exhibit a modest $A$ dependence \cite{benhar93}.  This makes them
an universal feature of atomic nuclei whose experimental determination
is of the utmost importance. The hadronic current operator
$J_{\mu}(q)$ in the above expression accounts for the coupling of the
electromagnetic field to the nuclear system.  In our calculations, the
one-body part ($J_{\mu}^{[1]}$) has been implemented in the standard
Impulse Approximation (IA) fashion, whereas the two-body currents
($J_{\mu}^{[2]}$) include the conventional pion-exchange and
intermediate $\Delta_{33}$-resonance terms \cite{jan97}.  The matrix
element of Eq.~(\ref{eq:transim}) contains 2,3,... A-body correlations
and is usually calculated with the aid of an expansion that is cut at
some order.  It is worth stressing that most of the cluster expansions
for correlated systems that are outlined in literature refer to
calculations addressing the ``inclusive'' response of the correlated
target nucleus to the electromagnetic probe.  Hereby, closure
properties can be exploited to help reducing the complexity of the
calculations.  In the $(e,e'p)$ case, for example, closure can be
applied when integrating over all excitation energies of the residual
$A-1$ system \cite{treleani}. This quantity, however, is
experimentally not accessible as it would involve integrations over
kinematics regions where the measured $(e,e'p)$ strength is
contaminated by pion production \cite{amparo}.  Here, we aim at
computing the effect of the correlations on some well defined parts of
the phase space in exclusive and semi-exclusive electronuclear
processes.  Accordingly, no closure relations can be applied and most of
the known cluster expansion techniques are not directly applicable.
In inclusive $^{12}$C$(e,e')$ \cite{giam} and $^4$He$(e,e'd)$
\cite{eed} calculations the lowest-order approximation (LOA) was
observed to account for the major $NN$ correlations effects.
Given that exclusive $A(e,e'NN)$ processes are confined to two-nucleon
phase space, three- and higher-body ``correlations'' are expected to
produce even smaller corrections than in the inclusive case.  For that
reason, the transition matrix element of Eq.~(\ref{eq:transim}) was
evaluated in the LOA. This procedure results in a clear separation
between the contributions from the uncorrelated mean-field wave
function that read
\begin{equation}
\left<\Psi_f  \right| \left( \sum _{i=1}^{A} J^{[1]}_{\mu}(i;q) + 
\sum _{i<j=1}^{A}J^{[2]}_{\mu}(i,j;q) \right)
\left| {\Psi_i} \right> \;,
\label{eq:meanfield}
\end{equation}
and the ones that can be unambiguously attributed to the $NN$
correlations
\begin{eqnarray}
& & \left<\Psi_f  \right|  \sum _{i < j=1} ^{A}  \Biggl[ 
\left( J^{[1]}_{\mu}(i;q) + J^{[1]}_{\mu}(j;q) + J^{[2]}_{\mu}(i,j;q) \right) 
\nonumber \\
& & \times \left( - g_c(r_{ij})+f_{t\tau}(r_{ij})\widehat{S_{ij}} \vec{\tau}_i
. \vec{\tau}_j \right) + h.c. \Biggr]
\left| {\Psi_i} \right> \;.
\label{eq:correlation}
\end{eqnarray}
The correlation functions $g_c$ and $f_{t\tau}$ establish the link
between the measurements and the nuclear many-body theories and
contain the information of the ``beyond mean-field'' structure of
nuclei.  In inclusive $A(e,e')$ and exclusive $A(e,e'p)$ reactions,
signals from the ``correlation'' part (\ref{eq:correlation}) in the
transition matrix elements are frequently obscured by the dominant
one-body term $J^{[1]}_{\mu}$ in Eq.~ (\ref{eq:meanfield}).  For
uncorrelated wave functions, two-nucleon knockout strength can solely
be generated by the one-body current through final state interaction
effects, e.g. $A(e,e'p)A-1$ followed by $A-1(p,p'n)$.  We will
consider kinematical regimes in which the contaminant effect of
rescattering processes is minimized. To that purpose we compute
proton neutron knockout processes from a light nucleus ($^{16}$O) with
the condition that the residual nucleus ($^{14}$N) is created at low
excitation energies. Moreover, we will consider the situation that the
two nucleons are ejected along the momentum transfer (``super-parallel
kinematics'').  Then, $A-1(p,p'n)$ rescattering processes involve
large momentum transfers and are heavily suppressed.  This feature of
heavily suppressed one-body current contributions to the uncorrelated
matrix element of Eq.~(\ref{eq:meanfield}), opens up good 
opportunities to acquire a precise understanding of $NN$ correlations
by studying two-nucleon knockout processes.


\begin{figure}
\begin{center}
\setlength{\unitlength}{1cm}
\begin{picture}(8,9)(0,0)
\put(-0.5,1.0){\mbox{\epsfysize=8.25cm\epsffile{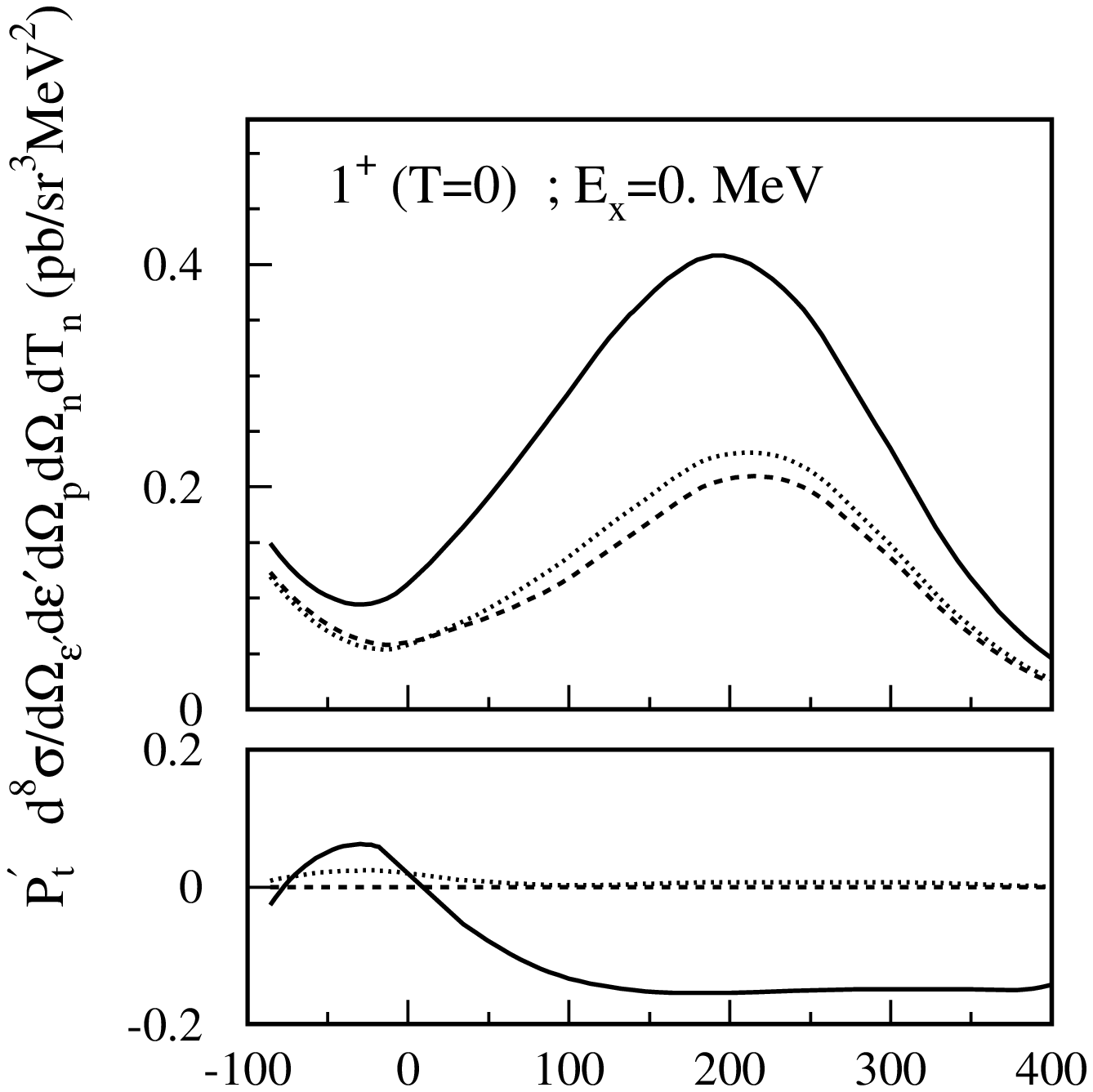}}}
\put(2.95,1.0){\mbox{\epsfysize=8.25cm\epsffile{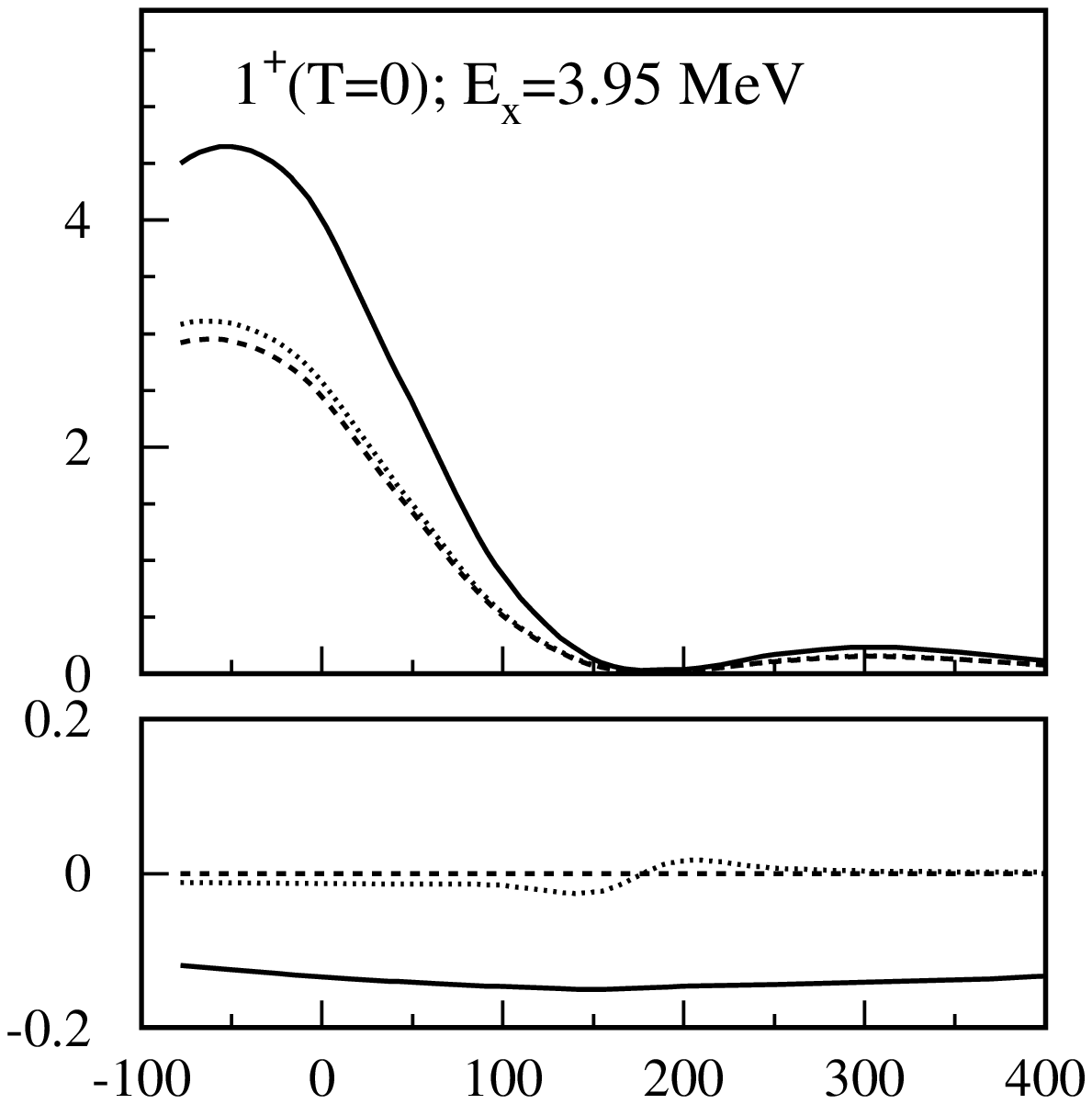}}}
\put(-0.50,-2.65){\mbox{\epsfysize=8.25cm\epsffile{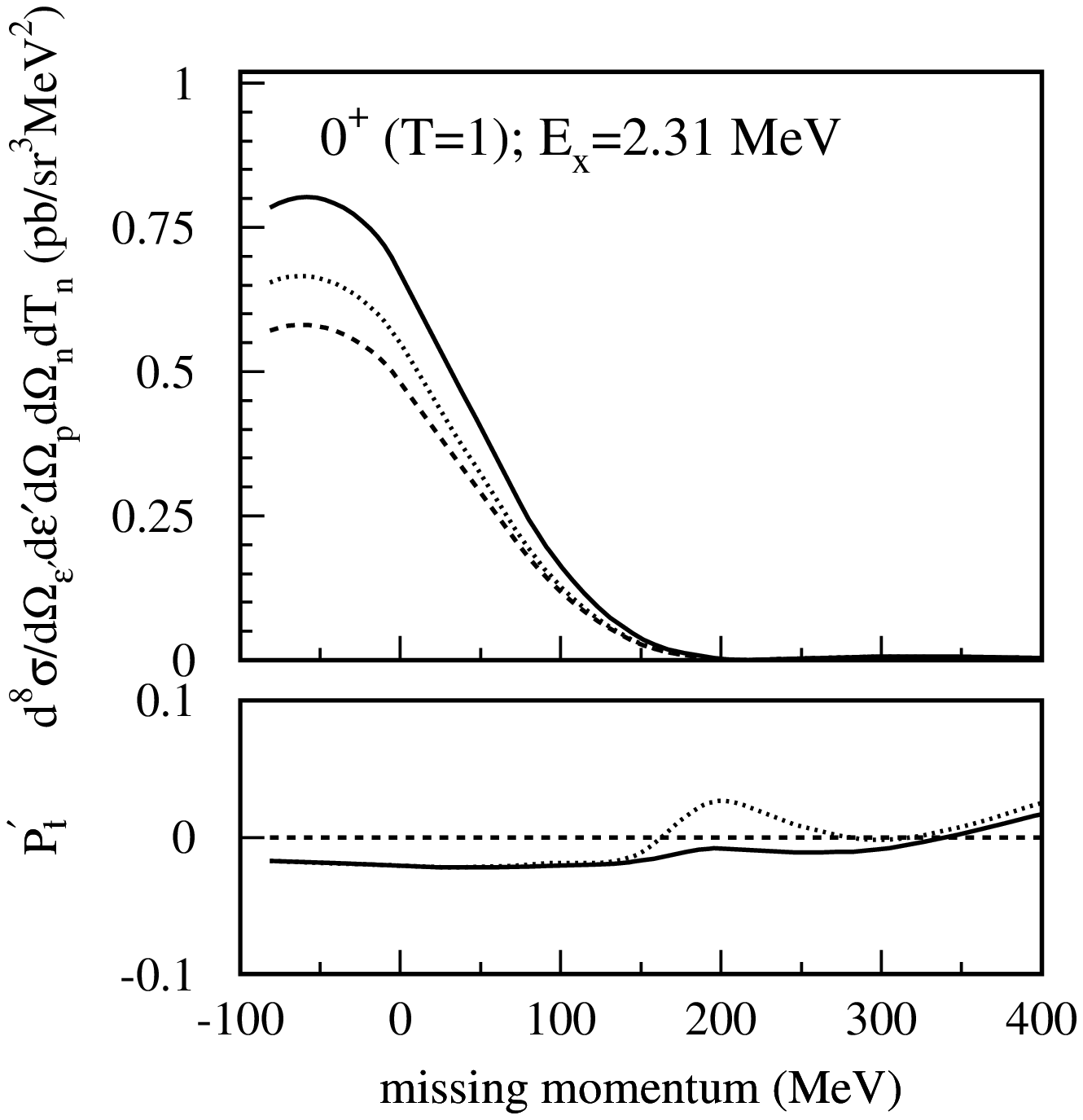}}}
\put(2.95,-2.65){\mbox{\epsfysize=8.25cm\epsffile{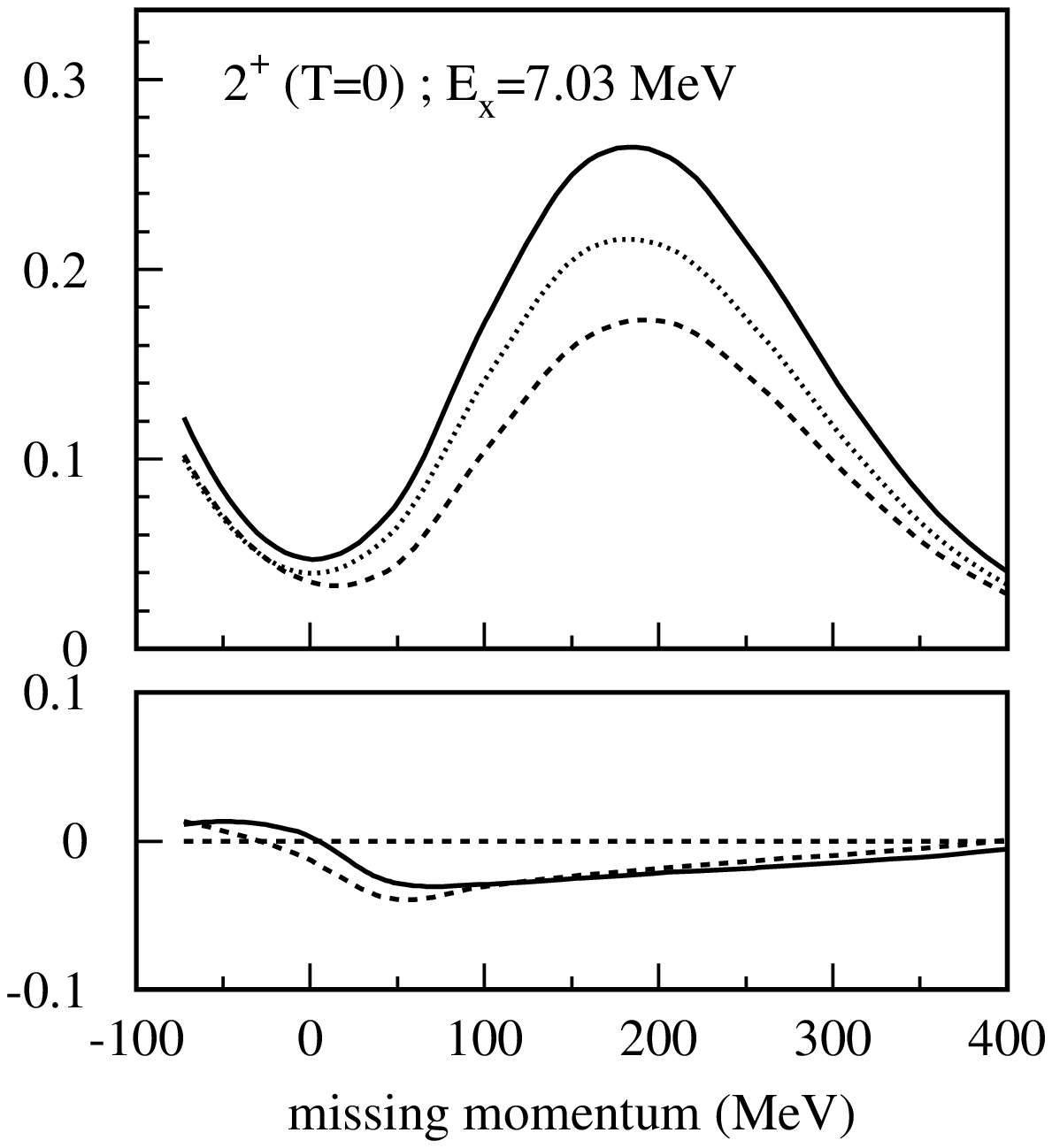}}}
\end{picture}
\end{center}
\caption{\em The missing momentum dependence of the $^{16}$O($e,e'pn$)
differential cross section and recoil polarization observable $P'_t$
for excitation of the low lying two-hole states in $^{14}$N at
$\epsilon$=855~MeV, $\epsilon '$=640~MeV and $\theta _{e'}$=18$^o$. The
calculations refer to the situation in which the proton (neutron) is
detected parallel (anti-parallel) to the direction of the momentum
transfer.  The solid curve is calculated in the distorted-wave
approximation including the two-body currents, central and tensor
correlations. The dashed line is the contribution from the
two-body currents.  The dotted lines omit tensor correlations from the
full calculation.}
\label{fig:eepn}
\end{figure}

Direct information on the precise role of tensor correlations in nuclei is expected
from triple coincidence $A(e,e'pn$) experiments, provided that one can
separate the signals from the electromagnetic coupling to a tensor
correlated proton-neutron pair from two-body current contributions.  
In $A(e,e'pp)$ and $A(e,e'pn)$ studies, a powerful tool in the
search for $NN$ correlations is the selectivity of the
final state with respect to the quantum numbers of the correlated
nucleon pair that actively participates in the reaction process
\cite{giusti98,giusti99,ryckebusch99}.  We have calculated
$^{16}$O($e,e'pn$)$^{14}$N cross sections for excitation of specific
states in $^{14}$N.  The results are displayed in Figure
\ref{fig:eepn} and include apart from the central and tensor
correlations, the ``uncorrelated'' contribution of competing
meson-exchange and intermediate $\Delta$-resonance two-body current
contributions.  The model employed to calculate these cross sections
includes a distorted wave description of the ejectiles and was
extensively discussed in Reference~\cite{jan97}. Restricting ourselves
to the (dominant) p-shell components, the following selectivity for
the quantum numbers of the active proton-neutron pair emerges when
considering $^{16}$O($e,e'pn$) decay to the different angular momentum
states in $^{14}$N
\begin{equation}
\begin{array}{ll}
J_R=0^+ & ^1S_0 (\Lambda=0), ^3P_1 (\Lambda=1)\\ J_R=1^+ & ^3S_1
(\Lambda=0,2), ^3P_{0,1,2} (\Lambda=1), \\ & ^1P_1 (\Lambda=1), ^3D_1
(\Lambda=0) \\ J_R=2^+ & ^1S_0 (\Lambda=2), ^3S_1 (\Lambda=0,2),
^3P_{1,2} (\Lambda=1), \\ & ^1D_2 (\Lambda=0), ^3D_2 (\Lambda=0)
\end{array}
\label{eq:moshin}
\end{equation}
where the standard convention ($^{2S+1}L_J$) for the relative
two-nucleon wave function is adopted and $\Lambda$ is the orbital
quantum number corresponding with the c.o.m. motion of the pair.  We
have calculated the differential cross section for these states in the
low-energy spectrum of $^{14}$N that are established to have a
two-hole character relative to the ground state of $^{16}$O
\cite{snelgrove69} and are therefore expected to be strongly populated
in a direct $^{16}$O$(e,e'pn)$ reaction \cite{giusti99}.  The two-body
overlap amplitudes employed in our calculations are from
Ref.~\cite{cohen70}.  The $g_c$ and $f_{t\tau}$ correlation functions
are those from \cite{pieper}.  They are obtained in variational
calculations for the $^{16}$O ground state with the Argonne $v_{14}$
NN potential.  We have selected so-called ``super-parallel''
kinematics which makes the two nucleons to move along the momentum
transfer.  It has become customary in the discussion of two-nucleon
knockout processes to present the results in terms of the pair missing
momentum. In a spectator model the pair missing momentum coincides
with the inital c.o.m. momentum of the ejected nucleon pair.  As
inferrred from the missing-momentum dependence of the cross sections
contained in Figure~\ref{fig:eepn}, the ground state has a mixed
$\Lambda$=0,2 character, whereas the $J_R=1^+$ state at $E_x=$3.95~MeV
reflects a $\Lambda$=0 shape. As evidenced by the results of
Figure~\ref{fig:eepn} the effect of the tensor correlations in
exclusive proton-neutron knockout is large.  Neglecting the tensor
correlations would result in a cross section for the ground-state
transition that is about a factor of two too small.  The strongest
sensitivity to the tensor correlations, though, is found in the peak
of the cross sections for the transition to the $1^+$ states at
respectively $E_x$=0 and 3.95~MeV.  As is clear from the pair
combinations contained in Eq.~(\ref{eq:moshin}) this enhanced
sensitivity to the tensor correlations corresponds with the situation
that the reaction is dominated by absorption on proton-neutron pairs
in a $^3S_1$ configuration. The predicted effect of the central
correlations is small in comparison with the tensor contributions.
For the transition to the $0^+$ and $2^+$ states, the effect of the
central and tensor correlations is about of equal importance and
relatively modest.  This can be explained by remarking that for these
states the $^3S_1$ configuration does not (0$^+$) or only marginally
contribute (2$^+$). A striking feature is the big difference in the
computed differential cross sections for the two 1$^+$ states.  A
detailed $^{16}$O$(\pi ^+,pp)^{14}$N measurement reported in
Ref.~\cite{schumacher98} determined that the transition cross section
has a mixed $\Lambda$=0,2 for the ground-state and a predominant
$\Lambda$=0 character for the $E_x$=3.95~MeV state. Moreover, the
differential cross section for the $E_x$=3.95~MeV state was determined
to be substantially larger than for the ground state.  In many
respects the $(\pi ^+,pp)$ process bears resemblance with the
$(e,e'pn)$ reaction and our computed $^{16}$O$(e,e'pn)$ differential
cross sections for the $1^+$ states exhibit exactly the same
qualitative features than those observed in the
$^{16}$O$(\pi^+,pp)^{14}$N data of Ref.~\cite{schumacher98}.

Also shown in Fig.~\ref{fig:eepn} are the predictions for the double
polarization observables $P_t'$. This variable can be determined in
$A(\vec{e},e'\vec{p}n)$ measurements and determines the recoil
polarization along the direction $\widehat{t}$ which is in the
reaction plane orthogonal to the direction of the ejected proton's
momentum.  In superparallel kinematics, $P_t'$ is uniquely determined
by the $W_{LT}'$ structure function.  The background of two-body
current contributions to proton-neutron knockout is almost exclusively
transverse.  With no ground-state correlations contributing, $P_t'$ is
extremely small. For these transitions with strong contributions from
the tensor correlations, the calculations predict large values $P_t'$,
though. In that respect, the recoil polarization observable is a
measure for the importance of tensor correlations in the medium.
Double polarization observables have been frequently shown to be
relatively free of ambiguities with respect to the final state
interaction \cite{ryckebusch99}.

\begin{figure}
\begin{center}
\mbox{\epsfxsize=9.cm\epsffile{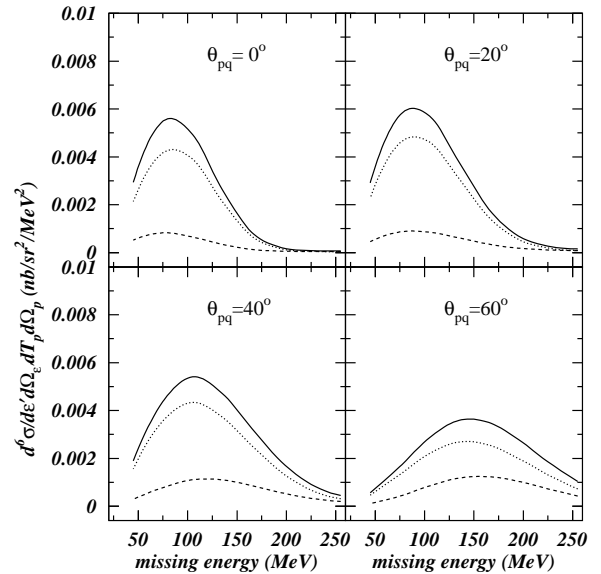}}
\end{center}
\caption{\em The missing-energy dependence of the computed contribution
from NN correlations to the semi-exclusive $^{16}$O($e,e'p$)
differential cross section at various proton polar angles.  Dashed
curves include only the central correlations, solid curves both
central and tensor correlations.  The dotted curve is the calculated
contribution from ($e,e'pn)$ including tensor and central
correlations. The electron kinematics is determined by
$\epsilon=1.2~GeV$, $\epsilon'=0.9~GeV$ and $\theta_e=16^o$.}
\label{fig:semiexclu}
\end{figure}

In general, triple coincidence measurements are challenging.  An
indirect way of accessing the ground-state correlations is the
semi-exclusive $A(e,e'p)$ process. Indeed, when probing higher missing
energies $E_m = \omega - T_p - T_{A-1}$, the residual nucleus is
created at a high excitation energy ($E_x \gtrsim $30~MeV).  In this
regime, exclusive single-nucleon knockout through the ``uncorrelated''
one-body current operator of Eq.~(\ref{eq:meanfield}) is heavily
suppressed and besides the mesonic degrees of freedom, the
``correlations'' terms in Eq.~(\ref{eq:correlation}) are expected to
feed the $A(e,e'p)$ channel.  As the ``correlations'' are
predominantly of two-body nature, the ``correlated'' $A(e,e'p)$
strength will manifest itself as two-nucleon knockout.  Often, in
interpreting semi-exclusive $A(e,e'p)$ and inclusive $A(e,e')$
processes a factorization scheme is adopted.  Hereby, the hadronic
part of the $A(e,e'p)$ cross section is written in terms of the
probability to find a nucleon with a certain momentum for a fixed
missing energy of the $A-1$ spectators (the so-called spectral
function).  For the results presented here, an unfactorized approach
is adopted.  The method was outlined in Ref.~\cite{jan97} and is based
on explicitly calculating the contribution from the operator
$\left(J^{[1]}_{\mu}(i;q) + J^{[1]}_{\mu}(j;q) \right) \left( -
g_c(r_{ij})+f_{t\tau}(r_{ij})\widehat{S_{ij}} \vec{\tau}_i
. \vec{\tau}_j \right)$ to the two-nucleon knockout channels
($A(e,e'pn$) and $A(e,e'pp$)) and integrating over the complete phase
space of the undetected nucleon.  The calculated missing energy
spectrum of the $^{16}$O$(e,e'p)$ reaction at $\omega$=300~MeV and
$\mid \vec{q} \mid$=416~MeV/c is shown in Figure~\ref{fig:semiexclu}
for a number of proton angles.  It is worth stressing that in the
absence of $NN$ correlations all computed strengths in
Figure~\ref{fig:semiexclu} would vanish identically. As the angle
$\theta_{pq}$, which is the polar angle of the detected proton
relative to the direction of the momentum transfer, grows higher
initial proton momenta are probed.  A striking feature of the results
is that the tensor correlations generate a few times more
``correlated'' $^{16}$O$(e,e'p)$ strength than the central short-range
correlations do.  At $\theta _{pq}$=40$^o$, where the missing (or,
initial proton) momentum ranges over $275 \leq p_m \leq 450$~MeV/c,
inclusion of the tensor correlations increases the strength with a
factor of five.  As clearly illustrated in Figure~\ref{fig:semiexclu},
most of the ``correlated'' $(e,e'p)$ strength can be ascribed to the
proton-neutron knockout channel.  With increasing polar angle $\theta
_{pq}$ higher missing momenta are probed and the central correlations
gain in relative importance.  This can be explained by considering
that central correlations refer to the hard part of the $NN$ force,
whereas tensor correlations are of somewhat larger ranges (1$\lesssim r
\lesssim$3~fm).  Qualitatively, the central and tensor correlations
exhibit a similar missing energy behaviour for all polar angles
included in Figure~\ref{fig:semiexclu}.  This feature reflects the
fact that both the central- and tensor correlations are mostly
affecting nucleon pairs in relative $S$ states, thus imposing
stringent kinematical constraints on the $(E_m,p_m)$ regions where the
strength attributed to the $NN$ correlations resides \cite{ciofi}.

Summarizing, we have presented a framework permitting a systematic
investigation of the effects of state-independent (central) and
state-dependent (tensor) correlations upon $A(e,e'NN)$ and
semi-exclusive $A(e,e'p)$ differential cross sections.  The
calculations reveal that the sensitivity of the $^{16}$O$(e,e'pn)$
cross sections to central correlations is rather modest.  On the
contrary, strong signals from the tensor correlations appear in these
regions of the $d\Omega _p dE_p d\Omega _n dE_n$ phase space whenever
the inital photoabsorption happens to occur on $^3S_1$ proton-neutron
pairs.  The tensor correlations are further predicted to produce
substantially more semi-exclusive $^{16}$O$(e,e'p)$ strength than the
central (Jastrow) correlations do.  This peculiar feature of how the
$NN$ correlations manifest themselves in electronuclear processes,
makes the proton-neutron knockout channel to dominate the
semi-exclusive $(e,e'p)$ strength and could be considered as a
microscopic confirmation of the suggested quasi-deuteron scaling of
the ``correlated'' $A(e,e')$ strength \cite{frankfurt}.  Concluding,
our results indicate that exclusive electronuclear reactions are a
promising tool for providing insight into the role of tensor
correlations in the nuclear medium.

This work was supported by the Fund for Scientific Research of
Flanders under Contract No 4.0061.99 and the University Research
Council.

\onecolumn
%

\end{document}